\documentclass{iau}

\usepackage{amsmath}
\usepackage{graphicx}
\usepackage{multirow}

\begin{document}

\lefttitle{Shin Toriumi}
\righttitle{Solar Sources of Flares and CMEs}

\jnlPage{1}{7}
\jnlDoiYr{2024}
\doival{10.1017/xxxxx}
\volno{388}
\pubYr{2024}
\journaltitle{Solar and Stellar Coronal Mass Ejections}

\aopheadtitle{Proceedings of the IAU Symposium}
\editors{N. Gopalswamy,  O. Malandraki, A. Vidotto \&  W. Manchester, eds.}

\title{Solar Sources of Flares and CMEs}

\author{Shin Toriumi}
\affiliation{Institute of Space and Astronautical Science, Japan Aerospace Exploration Agency, Sagamihara, Japan}

\begin{abstract}
Strong solar flares and coronal mass ejections (CMEs) are prone to originate within and near active regions (ARs) with a high magnetic complexity. Therefore, to better understand the generation mechanism of flares and the resultant CME eruption and to gain insight into their stellar counterparts, it is crucial to reveal how solar flare-productive ARs are generated and developed. In this review, first, we summarize some general aspects of solar flares and key observational characteristics of such ARs. Then, we discuss a series of flux emergence simulations that were performed to elucidate the subsurface origins of their complexity and introduce state-of-the-art models that consider the effect of turbulent thermal convection. Future flare observations using SOLAR-C, a next-generation high-throughput extreme ultraviolet spectroscopy mission, are also discussed.
\end{abstract}


\maketitle

\section{Introduction}

\begin{figure}
\begin{center}
\includegraphics[width=0.95\textwidth]{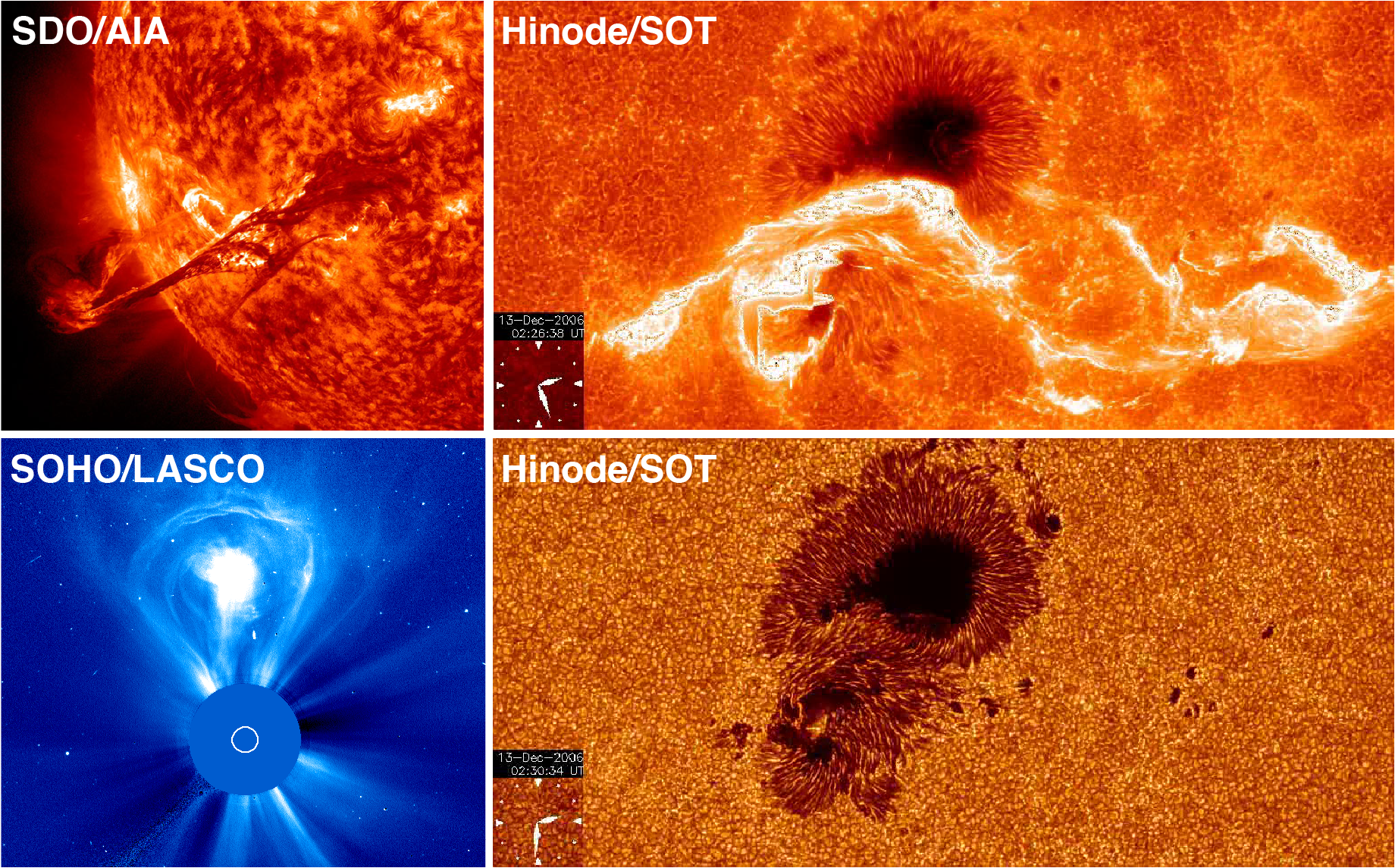}
\caption{Collection of solar flare and CME observations. Clockwise from top left: Filament eruption captured by SDO/AIA (credit: NASA), flare ribbons and sunspots captured by Hinode/SOT (NAOJ/JAXA), and a light-bulb-shaped CME captured by SOHO/LASCO (NASA/ESA).}
\label{fig:flare}
\end{center}
\end{figure}

Solar flares, especially strong ones, which may develop into coronal mass ejections (CMEs), are known to emanate from active regions (ARs) (see Figure \ref{fig:flare}) \citep{2019LRSP...16....3T}. Recent stellar observations have shown that even more massive events occur in late-type dwarf stars \citep[for example,][]{2012Natur.485..478M}. However, with the current telescope capabilities, it is difficult to observe flare-hosting starspots in spatially resolved images. Therefore, in order to understand the solar and stellar magnetic activities common to these stars, it is important to deepen our knowledge of the solar ARs that produce flares and CMEs. Accordingly, this review summarizes the key observational features of flare-productive ARs and numerical modeling approaches to understand their formation mechanisms.

The remainder of this paper is organized as follows. Section \ref{sec:observations} presents key aspects and observational features of flare-productive ARs. Sections \ref{sec:simulations} and \ref{sec:r2d2} present the numerical modeling. In Section \ref{sec:solarc}, we discuss future flare observations with the introduction of the next-generation mission SOLAR-C. Finally, Section \ref{sec:summary} summarizes this review and provides perspectives for future studies.

\section{Observations}\label{sec:observations}

In astronomy, the term {\it flare} describes the sudden amplification of electromagnetic waves over various wavelength ranges. Solar flares may be observed as brightening in the X-ray, ultraviolet (UV), visible, infrared, and radio bands. The energy released in a single event ranges from $10^{29}\ {\rm erg}$ to $10^{32}\ {\rm erg}$, with time scales ranging from tens of minutes to hours and size scales ranging in the order of $10\ {\rm Mm}$ to $100\ {\rm Mm}$.

\begin{figure}
\begin{center}
\includegraphics[width=0.8\textwidth]{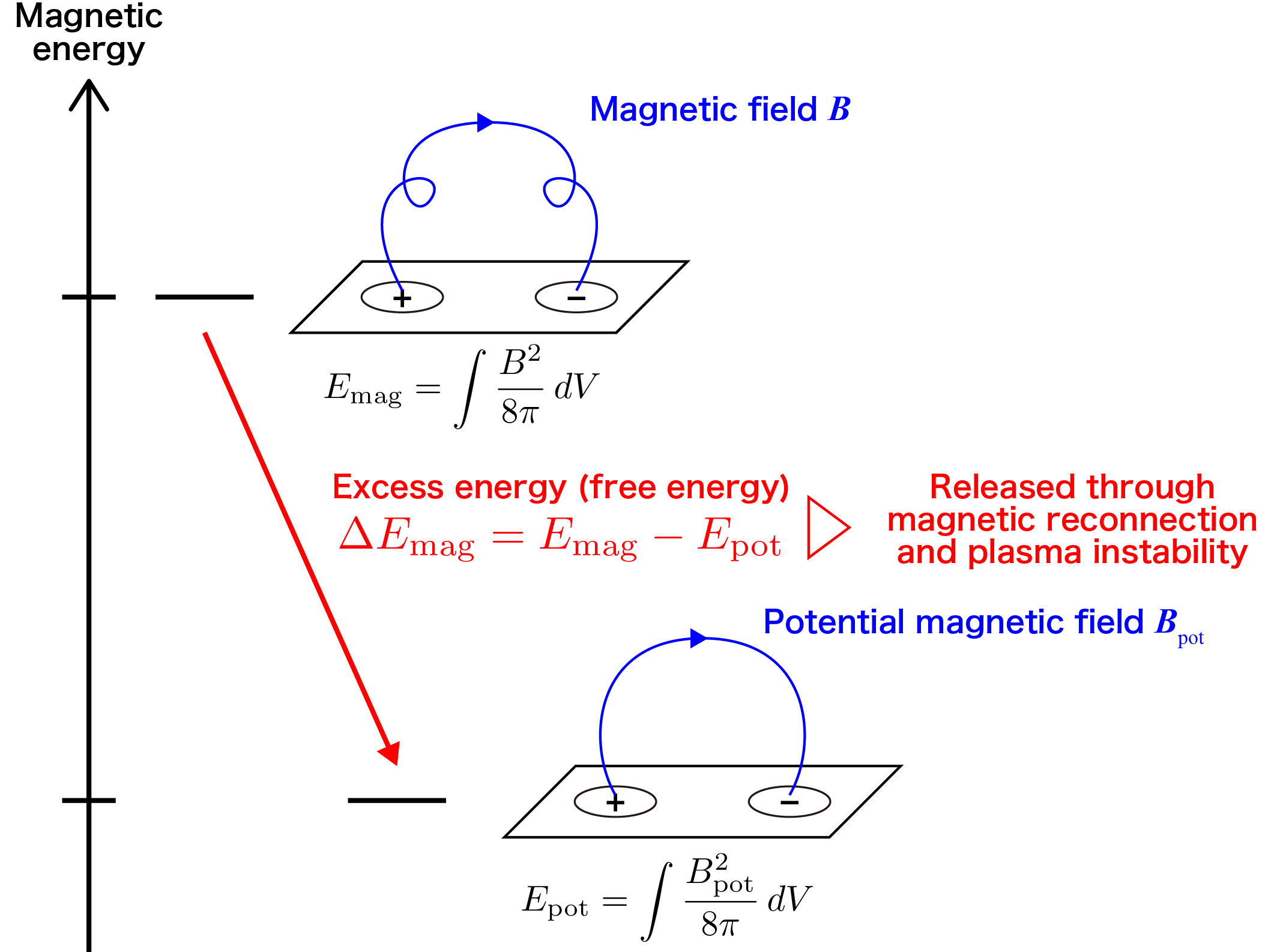}
\caption{Energetics of solar flares. The differential energy between the two levels, the free energy $\Delta E_{\rm mag}$, is released through the flare. Note that only a fraction of the free energy is actually released in each event.}
\label{fig:energy}
\end{center}
\end{figure}

From an energetics perspective, solar flares can be viewed as a phenomenon that releases excess magnetic energy, or free energy, which accumulates in the coronal magnetic field. Figure \ref{fig:energy} shows the energy diagram for the pre-flare magnetic field $\mathbf{B}$ and the potential magnetic field $\mathbf{B}_{\rm pot}$, which is the minimum energy state. The free energy can be described as
\begin{eqnarray}
  \Delta E_{\rm mag} = E_{\rm mag}-E_{\rm pot}
  = \int \frac{B^{2}}{8\pi}\, dV - \int \frac{B^{2}_{\rm pot}}{8\pi}\, dV.
\end{eqnarray}
It is released through magnetic reconnection and plasma instability during a flare event, although only a (small) fraction of the stored free energy is released during a single event. During the flare, free energy is converted into thermal energy (hot plasma above 1 MK), kinetic energy (i.e., CME), and particle acceleration (solar energetic particles). This suggests that the magnetic field must be non-potential for a flare to occur.

A key observational feature of flare-productive ARs is the sheared polarity inversion line (PIL). Such PILs have been shown to have a strong horizontal field ($>4000\ {\rm G}$), a strong $B_{z}$ gradient ($\sim 100\ {\rm G}\ {\rm Mm}^{-1}$), and a strong magnetic shear ($80^{\circ}$--$90^{\circ}$) \citep[e.g.,][]{1958IzKry..20...22S,1984SoPh...91..115H,2007ApJ...655L.117S}, and are sometimes accompanied by sunspot rotation \citep{2009SoPh..258..203M,2003SoPh..216...79B}. Helical magnetic structures or flux ropes can often be observed above the PILs before an eruption. They are seen as filaments in H$\alpha$ and other chromospheric lines and as sigmoids in soft X-rays and UV lines \citep{1996ApJ...464L.199R,1999GeoRL..26..627C,2002SoPh..207..111P}. When it erupts, the flux rope evolves into the magnetic skeleton of the CME. All the above-mentioned structures indicate that the magnetic field is non-potential.

A representative statistical study on flaring ARs is \cite{2000ApJ...540..583S}, where it was shown that there is a positive correlation between the spot area and maximum flare magnitude in terms of soft X-ray flux.\footnote{It is pointed out by \citet{1998ApJ...508..885F}, \citet{2003ApJ...598.1387P}, \citet{2022ApJ...927..179T}, and other studies that the quiescent X-ray flux also correlates with the AR area (or the total magnetic flux).} Thus, one of the key conditions is the AR area, and this tendency may indicate that larger ARs can hold larger amounts of magnetic energy.

\citet{2000ApJ...540..583S} also showed that the difference in sunspot configuration is also an important factor. Compared to the $\beta$-spots, which have a simple bipolar structure, $\delta$-spots, in which umbrae of positive and negative polarities reside in a common penumbra, are prone to produce massive flares \citep{1960AN....285..271K}. A number of statistical studies \citep[e.g.,][]{1985SoPh...96..293M,2002SoPh..209..361T,2017ApJ...834...56T} support this finding. The importance of $\delta$-spots as flare-prolific ARs is most evident in NOAA AR 5395, which was a typical $\delta$-spot with numerous positive and negative umbrae within a massive penumbra \citep{1991ApJ...380..282W}. This AR produced more than 200 flares, including the X15-class event, and was responsible for the magnetic storm that caused a blackout in Quebec and damaged a transformer in New Jersey in March 1989. Therefore, the magnetic and morphological complexity of ARs is another indication, which is natural because the free energy accumulates in a non-potential magnetic field.

In addition, the rapid evolution of an AR is also an important factor, as magnetic energy must accumulate sufficiently faster than it dissipates \citep{2017RNAAS...1...24S,2022ApJ...938..117N}. The key elements of flare-producing ARs are summarized as follows:
\begin{itemize}
\item Area: The AR has a large amount of magnetic energy.
\item Complexity: The AR has a large non-potentiality, i.e., large free energy.
\item Rapid evolution: The accumulation of magnetic energy is sufficiently faster than its dissipation.
\end{itemize}

How do these complexly shaped ARs form? ARs are formed by the emergence of magnetic flux from the deep convection zone; then, what happens down there? It is impossible to observe the convection zone via direct optical observation, and helioseismology may not be sensitive enough to detect emerging fluxes with high spatial and temporal resolutions \citep[for example, see][]{2009SSRv..144..175K}.

\section{Simulations}\label{sec:simulations}

One way to overcome this problem and understand the flux emergence from the deep convection zone is to reproduce the flux emergence using magnetohydrodynamic simulations. In a typical simulation, a 2D or 3D computational domain is prepared, which is gravitationally stratified and consists of a convection zone, photosphere, chromosphere, and corona. Initially, at $t=0$, a magnetic flux tube is placed in the convection zone, and buoyant emergence (Parker instability) is induced, for example, by adding a velocity perturbation or reducing the density \citep{1989ApJ...345..584S,2001ApJ...554L.111F,2004A&A...426.1047A}. Many attempts have been made to model complex-shaped (i.e., flare-producing) ARs by controlling the conditions of the initial flux tubes.

The first flaring AR formation scenario is the kink instability of the magnetic flux tubes. If the twist of a flux tube is sufficiently strong, kink instability is triggered and the flux tube deforms, forming two strongly rotating sunspots and complex magnetic structures in the photosphere. A theoretical analysis of the kink instability of the emerging flux was provided by \cite{1996ApJ...469..954L}, and it was first modeled by \cite{1998ApJ...505L..59F} and \cite{1999ApJ...521..460F}. Subsequent simulations have shown that kink instability forms complex magnetic structures reminiscent of $\delta$-spots \citep{2015ApJ...813..112T,2017ApJ...850...39T,2018ApJ...864...89K}.

\cite{2014SoPh..289.3351T} and \cite{2015ApJ...806...79F} proposed another possibility in which a single magnetic flux tube rises at two locations. In this case, because the flux tube produces a double-arched (i.e., M-shaped) loop instead of a classical $\Omega$-loop, two pairs of bipoles are formed in the photosphere. The positive and negative spots collide at the center of the domain, forming a densely packed $\delta$-spot.

The third concept is the emergence of multiple magnetic flux tubes intertwined with each other. From the sunspot record over two solar cycles, \cite{2016ApJ...820L..11J} found that the fraction of $\gamma$- and/or $\delta$-spots increases during the solar maximum, and suggested the possibility that the pileup of emerging magnetic fluxes produces complex-shaped ARs. Simulation results by \citet{2018ApJ...857...83J} showed that when magnetic flux tubes collide below the photosphere, depending on their mutual angle, the structure of the photospheric magnetic fields becomes complex.

However, these numerical simulations were highly idealized. In some models, the subsurface layer was just an adiabatically-stratified atmosphere; thus, no convection occurred. Even in the convective-flux-emergence simulations, flux tubes were kinematically inserted from the shallow bottom boundary for computational constraints \citep{2010ApJ...720..233C,2014ApJ...785...90R}. Thus, it remains unclear how deep large-scale thermal convection affects flux emergence.

\section{Recent Attempts: R2D2 Simulations}\label{sec:r2d2}

To overcome this difficulty and reproduce realistic thermal convection in a 3D computational box that spans the entire convection zone, the Radiation and Reduced speed of sound technique (RSST) for Deep Dynamics, or R2D2, simulation code was developed by \cite{2019SciA....5.2307H}. R2D2 solves the magnetohydrodynamic equations with realistic radiation transfer and the equation of state, and implements RSST, enabling the reproduction of an inserted flux tube forming an AR in a self-consistent manner as a result of interaction with convective upflows/downflows in a deep domain (down to $-200$ Mm). Using R2D2, \cite{2019ApJ...886L..21T} and \cite{2020MNRAS.498.2925H} simulated $\delta$-spot formation via the collision of positive and negative polarities, as a single twisted flux tube emerges at two locations. The $\delta$-spot is formed by two co-rotating sunspots, creating a sheared magnetic arcade or a flux rope-like structure above the PIL. This supports the concept of idealized simulations proposed by \cite{2014SoPh..289.3351T} and \cite{2015ApJ...806...79F} (Section \ref{sec:simulations}) and indicates that such situations could occur in the actual Sun. In addition, \cite{2022MNRAS.517.2775K} found that the success or failure of emergence depends significantly on the initial position of the magnetic flux tube relative to the background convection.

\begin{figure}
\begin{center}
\includegraphics[width=\textwidth]{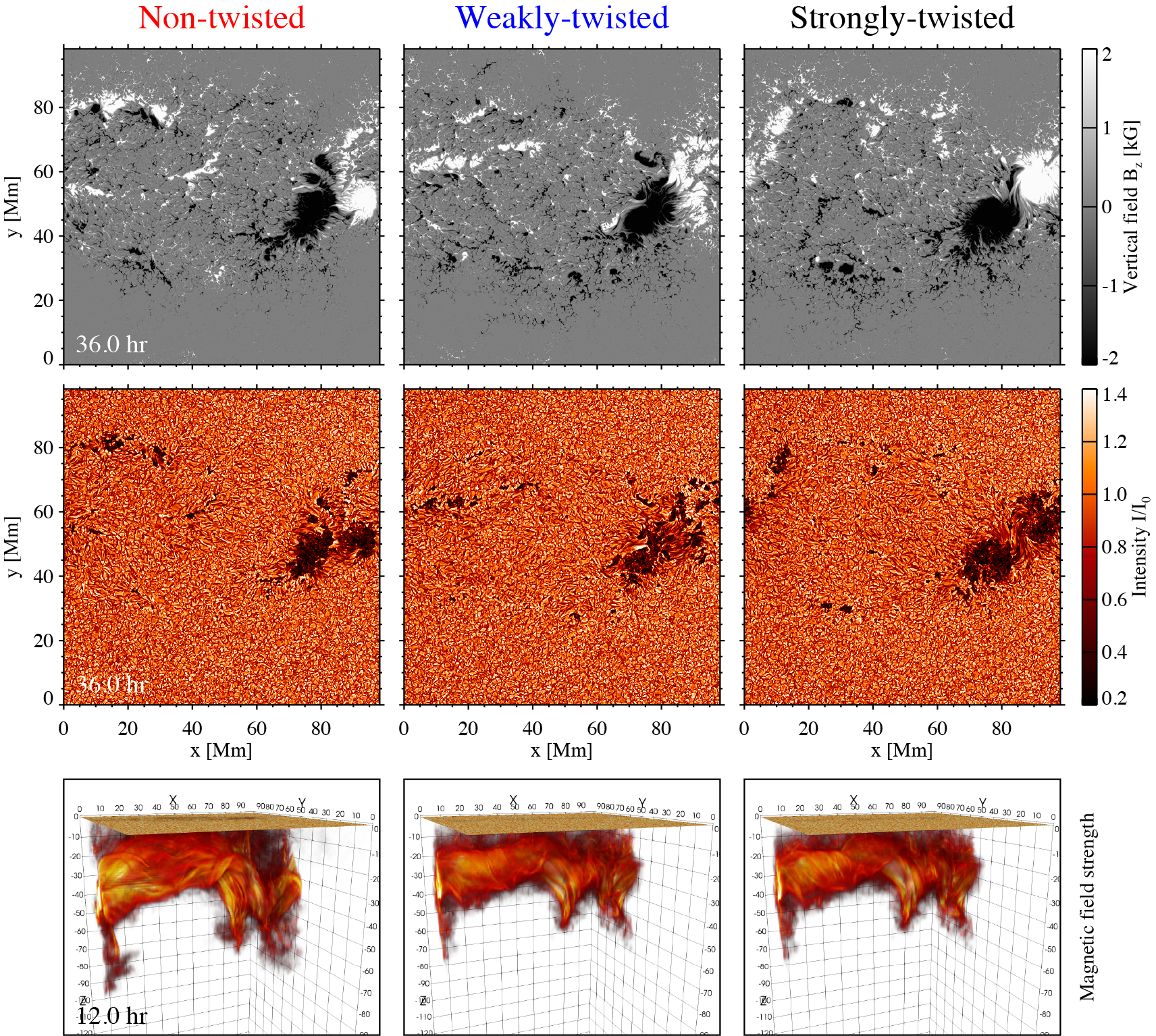}
\caption{(Top) Vertical magnetic field strength ($B_{z}$) in the photosphere, (Middle) emergent intensity, and (Bottom) 3D volume rendering of the total magnetic field ($|\mathbf{B}|$) for the non-twisted ($q/q_{\rm cr}=0$), weakly-twisted ($q/q_{\rm cr}=1/4$), and strongly-twisted ($q/q_{\rm cr}=1/2$) flux tube cases. Image reproduced with permission from \cite{2023NatSR..13.8994T}, copyright by the authors.}
\label{fig:tile3}
\end{center}
\end{figure}

Recently, \cite{2023NatSR..13.8994T} investigated the difference in flux emergence by varying only the initial twist of the flux tubes while keeping the background convection identical for the three cases (Figure \ref{fig:tile3}). The twist strength $q/q_{\rm cr}$ was varied in three cases: $q/q_{\rm cr}=[0, 1/4, 1/2]$, where $q_{\rm cr}$ is the critical twist strength for kink instability \citep{1996ApJ...469..954L}. Consequently, in all cases, the flux tube reached the photosphere and formed a bipolar AR. After the positive spot crossed the horizontal boundary (the $x=0$ boundary in the top panels of Figure \ref{fig:tile3}), the positive and negative spots collided head-on around $x=90\ {\rm Mm}$ owing to the periodic boundary condition, and eventually formed a $\delta$-spot. There was little difference in the amount of magnetic flux appearing in the photosphere for the three cases. However, the magnetic helicity measured in the photosphere was greater for $q/q_{\rm cr}=1/2$ than that for $q/q_{\rm cr}=1/4$. The amount of injected relative magnetic helicity was also estimated by measuring the relative helicity flux through the photosphere. Interestingly, even in the un-twisted case (i.e., $q/q_{\rm cr}=0$), a finite amount of magnetic helicity was detected, which explains the medium-sized solar flares. Because this simulation did not consider solar rotation, the finite injection of magnetic helicity was purely an effect of background convection.

A detailed analysis revealed that background convection rotates the magnetic flux below the sunspots and provides helicity; as the sunspots develop, magnetic fields extend along the downflow plumes in the convection zone. The external plasma flows into the plumes, creating local eddies that rotate the magnetic fields below the sunspots. In the photosphere, spot rotation occurs and magnetic helicity is injected into the upper atmosphere. This result indicates that it is possible to produce medium-sized flares based purely on the effect of background turbulence.

\section{Flare Observations with SOLAR-C}\label{sec:solarc}

\begin{figure}
\begin{center}
\includegraphics[width=0.75\textwidth]{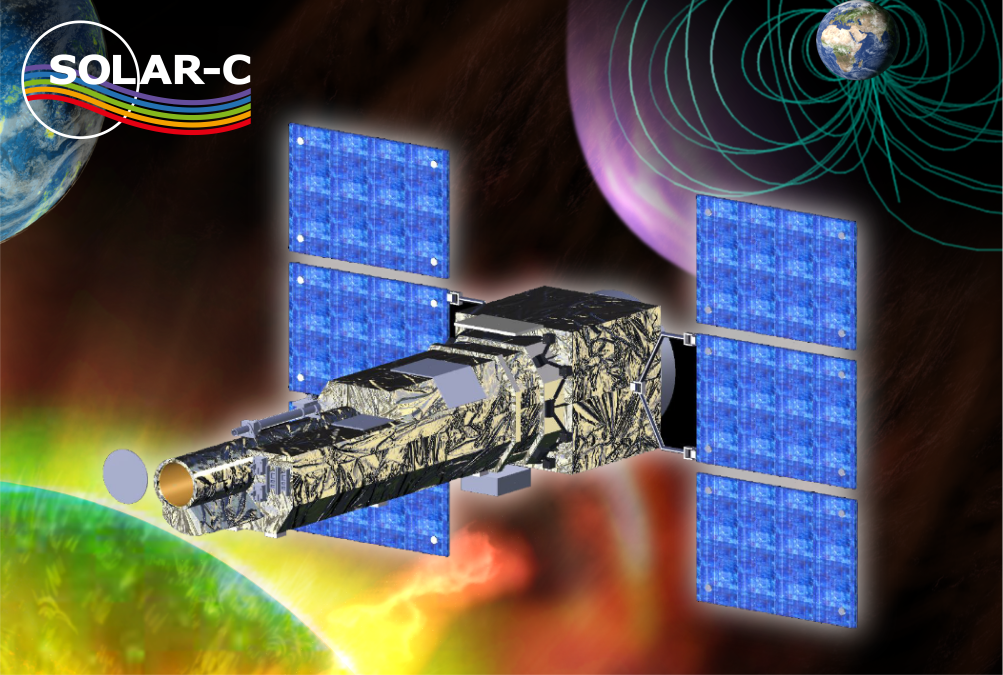}
\caption{SOLAR-C with its onboard telescope EUVST will unravel the formation mechanisms of the hot plasma and the Sun's effects on Earth and the solar system ultimately seeking the origin of the solar system and life (Credit: JAXA/NAOJ).}
\label{fig:solarc}
\end{center}
\end{figure}

This section focuses on the introduction to the solar flare observations to be performed by JAXA's next-generation solar-observing satellite, SOLAR-C (Figure \ref{fig:solarc}), which is scheduled to be launched in 2028 \citep{2020SPIE11444E..0NS}. SOLAR-C will aim at two science objectives, namely, it will aim to
\begin{enumerate}
\item[I.] Understand how fundamental processes lead to the formation of the solar atmosphere and the solar wind; and
\item[II.] Understand how the solar atmosphere becomes unstable, releasing the energy that drives solar flares and eruptions.
\end{enumerate}
SOLAR-C is equipped with the EUV High-throughput Spectroscopic Telescope (EUVST) onboard, whose three unique aspects are as follows:
\begin{enumerate}
\item Seamless and broad temperature coverage: from $\log{T}=4$ to $7$;
\item High spatial/temporal resolution: $0.4$ arcsec, $0.5$-s exposure; and
\item High dispersion spectroscopy: velocity resolution of $2\ {\rm km\ s}^{-1}$.
\end{enumerate}

Magnetic reconnection is thought to be an important process in solar flares. However, it has been difficult to reveal these processes using the capabilities of previous telescopes. For example, the magnetic reconnection zone is much darker than the surrounding area, and the resolution of the telescopes is insufficient. By taking advantage of spectroscopy with high spatial and temporal resolutions, SOLAR-C will reveal how solar flares occur. In particular, it answers the questions of how magnetic reconnection occurs fast enough to explain the observed flares and how the magnetic energy is stored and suddenly released. Repeated scanning of the entire AR over days also reveals how energy accumulates in the AR and how local changes destabilize the entire magnetic system.

The spectral coverage of EUVST is 17--21.5 nm (SW) and 46--122 nm (LW), which includes a variety of EUV lines that are sensitive to the chromospheric, transition-region, coronal, and flare temperature plasmas (0.02--15 MK). The field of view is designed as $280\times 280\ {\rm arcsec}^{2}$, which is sufficient to cover regular-sized ARs.

SOLAR-C is an international mission with the participation of the USA (NASA) and European countries (ESA and the space agencies of Germany, France, Italy, Switzerland, and Belgium). These institutions will build the detectors, fabricate the grating, and lead the integration and verification of the EUVST components. The SOLAR-C project also plans the coordinated observation campaigns with NASA's MUSE mission \citep[][target launch in 2027]{2020ApJ...888....3D} and other ground-based facilities to follow the energy transport from the photosphere to the corona, which allows for detailed investigations of the occurrence mechanisms of solar flares and CMEs.

\section{Summary and Perspectives}\label{sec:summary}

A series of flare observations have revealed that the characteristics of flare-prolific ARs can be summarized as area, complexity, and rapid evolution, which correspond to the amount of magnetic energy to be stored, the non-potentiality of the magnetic field, and fast energy storage, respectively. One future direction for exploration is to determine whether these features are also observed in starspots. Statistical studies have been conducted on starspot areas \citep[for example,][]{2013ApJ...771..127N}. However, is it possible to obtain information on complexity?

Flux emergence simulations revealed that the strong twist of the emerging flux and the interaction between the fluxes contributed to AR complexity. However, these simulations are highly ideal, and realistic modeling that includes convection has become available in recent years, as represented by the R2D2 code. It was found that background convection alone can supply sufficient magnetic helicity to explain medium-sized solar flares. Future starspot modeling would also be of interest.

In addition, for future observational studies, the SOLAR-C satellite with a new ultraviolet spectrometer, the EUVST, is scheduled to be launched in 2028. SOLAR-C will cooperate with other solar-observing satellites and ground-based telescopes to reveal the energy storage in ARs and the energy releasing during flares. Collaboration with stellar XUV studies should also be promoted.

\section*{Acknowledgments}
The author wishes to thank the organizers of the IAU Symposium No. 388 Solar and Stellar Coronal Mass Ejections for inviting him to the symposium. This work was supported by JSPS KAKENHI Grant Nos. JP20KK0072 (PI: S. Toriumi), JP21H01124 (PI: T. Yokoyama), and JP21H04492 (PI: K. Kusano).

\section*{Discussion}

{\it Jie Zhang:} Thanks for a very nice review talk. In your opinion or based on your experience, what is the dominant mechanism that produces twists in the corona leading to solar eruption? To be more specific, is the twist mainly from the bodily emergence of twisted flux tubes from the sub-photosphere, or is it mainly from the cancellation/shearing motion on the photospheric surface?

{\it Shin Toriumi:} Thank you for your positive comments and interesting question. I think that both mechanisms are possible. However, my simulation results show that while the twist of the flux tube itself is important, the external flow field of thermal convection, which probably causes cancellation/shearing in the photosphere, makes a non-negligible, or perhaps significant, contribution.

{\it Andrei Zhukov:} You rightly placed emphasis on $\delta$-configurations, as we know from observations that they are very flare-prolific. However, I would like to understand why is it the $\delta$ that is so important? Why is it important that the two strong magnetic polarities are engulfed by the same penumbra? Why does the penumbra matter in comparison with the situation of two strong magnetic polarities adjacent to each other but without a common penumbra?

{\it Shin Toriumi:} It is just empirically known that $\delta$-spots produce strong events. However, there may be several reasons for this. From my perspective, if an AR has $\delta$-spots, the chances that the positive and negative polarities are very close or that the penumbra is large (i.e., the AR harbors a large amount of free energy) increase.

{\it Xudong Sun:} You mentioned the possibility to try realistic flux emergence models for starspots. What would be the targets?

{\it Shin Toriumi:} The easiest task would be to make the magnetic flux tubes thicker in order to create starspots with larger amounts of magnetic flux. It would also be interesting to attempt flux emergence simulations for non-G-type stars by changing the background stratification profile.

\end{document}